\documentclass[journal]{IEEEtran}
\pdfoutput=1
\usepackage{graphicx}
\usepackage{array}
\usepackage{placeins}
\usepackage{soul}
\usepackage{titling}

\ifCLASSINFOpdf

\else

\fi

\usepackage{xcolor}
\usepackage{paralist}
\usepackage{multirow}

\usepackage{amsmath}
\usepackage{fancyhdr}
\fancypagestyle{firstpage}{
\lfoot[\footnotesize{\linewidth}]{\small{ \textcopyright 2020 IEEE.  Personal use of this material is permitted.  Permission from IEEE must be obtained for all other uses, in any current or future media, including reprinting/republishing this material for advertising or promotional purposes, creating new collective works, for resale or redistribution to servers or lists, or reuse of any copyrighted component of this work in other works.}}
}

\title{Deep Neural Network for Respiratory Sound Classification in Wearable Devices Enabled by Patient Specific Model Tuning}

\author{Jyotibdha Acharya$^\dag$\thanks{$^\dag$ Jyotibdha Acharya is with HealthTech NTU, Interdisciplinary Graduate Program, Nanyang Technological University, Singapore.}, Student Member, IEEE and Arindam Basu$^{\dag\dag}$\thanks{$^{\dag\dag}$Arindam Basu is with School of Electrical and Electronic Engineering, Nanyang Technological University, Singapore.}, Senior Member, IEEE}
\date{}

\begin{document}

\maketitle

\thispagestyle{firstpage}
\setcounter{page}{1}
\begin{abstract}
The primary objective of this paper is to build classification models and strategies to identify breathing sound anomalies (wheeze, crackle) for automated diagnosis of respiratory and pulmonary diseases. In this work we propose a deep CNN-RNN model that classifies respiratory sounds based on Mel-spectrograms. We also implement a patient specific model tuning strategy that first screens respiratory patients and then builds patient specific classification models using limited patient data for reliable anomaly detection. Moreover, we devise a local log quantization strategy for model weights to reduce the memory footprint for deployment in memory constrained systems such as wearable devices. The proposed hybrid CNN-RNN model achieves a score of $66.31\%$ on four-class classification of breathing cycles for ICBHI'17  scientific challenge respiratory sound database. When the model is re-trained with patient specific data, it produces a score of $71.81\%$ for leave-one-out validation. The proposed weight quantization technique achieves $\approx 4 \times$ reduction in total memory cost without loss of performance. The main contribution of the paper is as follows: Firstly, the proposed model is able to achieve state of the art score on the ICBHI'17 dataset. Secondly, deep learning models are shown to successfully learn domain specific knowledge when pre-trained with breathing data and produce significantly superior performance compared to generalized models. Finally, local log quantization of trained weights is shown to be able to reduce the memory requirement significantly. This type of patient-specific re-training strategy can be very useful in developing reliable long-term automated patient monitoring systems particularly in wearable healthcare solutions.
\end{abstract}


\IEEEpeerreviewmaketitle

\section{Introduction}
\label{intro}
Two most clinically significant lung sound anomalies are wheeze and crackle. Wheeze is a continuous high pitched adventitious sound that results from obstruction of breathing airway. While normal breathing sounds have majority of their energy concentrated in 80-1600Hz~\cite{gavriely1984measurement}, wheeze sounds have been shown to be present in the frequency range 100Hz-2KHz. Wheeze is normally associated with patients suffering from asthma, chronic obstructive pulmonary disease (COPD) etc. Crackles are explosive and discontinuous sounds present during inspiratory and expiratory parts of breathing cycle with a significantly smaller duration compared to the total breathing cycle. Crackles have been associated with obstructive airway diseases and interstitial lung diseases~\cite{piirila1995crackles}.

Auscultation has been used historically for screening and monitoring respiratory diseases. It provides a simple and non-invasive approach to detect respiratory and cardiovascular diseases based on lung sound abnormalities. But these methods suffer from two disadvantages. Firstly, a trained medical professional is required to diagnose a patient based on adventitious lung sounds and therefore, disproportionate number of medical practitioners compared to overall population hinders the speed at which patients are tested. Secondly, even if the patients are diagnosed by experienced professionals, there might be subjectivity in the diagnosis due to dissimilar interpretation of the respiratory sounds by different medical professionals~\cite{bahoura2004respiratory}.

So, in the past decade several attempts were made to design algorithms and feature extraction techniques for automated detection of breathing anomalies. Some popular feature extraction techniques used include spectrogram~\cite{acharya2017feature}, Mel-Frequency Cepstral Coefficients (MFCC)~\cite{lin2016automatic}, wavelet coefficients~\cite{bahoura2009pattern}, entropy based features~\cite{zhang2009novel} etc. Several machine learning (ML) algorithms have been developed in past few years to detect breathing sound anomalies such as  logistic regression~\cite{bokov2016wheezing}, Dynamic Time Wrap (DTW), Gaussian mixture model (GMM)~\cite{sen2015comparison}, random forest~\cite{acharya2017feature}, Hidden Markov Model (HMM)~\cite{jakovljevic2018hidden} etc. An exploration of existing literature reveals some conspicuous issues with these approaches. Firstly, most of the ML algorithms use manually crafted highly complex features suitable for their algorithms and due to absence of publicly available datasets, it was hard to compare the efficacy of the feature extraction methods and algorithms proposed~\cite{pramono2017automatic}. Secondly, most of the strategies were developed for a binary classification problem to identify either wheeze or crackle and therefore, not suitable for multi-class classification to detect wheeze and crackle simultaneously~\cite{chen2019triple}. These drawbacks make these approaches difficult to apply in real world scenarios.

Deep learning has gained a lot of attention in recent years due to its unparalleled success in a variety of applications including clinical diagnostics and biomedical engineering~\cite{litjens2017survey}. A significant advantage of these deep learning paradigms is that there is no need to manually craft features from the data since the network learns useful features and abstract representations from the data through training. Due to wide success of convolutional neural networks (CNN) in image related tasks, they have been extensively used in biomedical research for image classification~\cite{hosseini2016alzheimer}, anomaly detection~\cite{van2016fast}, image segmentation~\cite{song2015accurate}, image enhancement~\cite{oktay2016multi}, automated report generation~\cite{kisilev2016medical} etc. There have been multiple successful applications of deep CNNs in diagnosis of cardiac diseases~\cite{tran2016fully}, neurological diseases~\cite{van2017deep}, cancer~\cite{kooi2017discriminating} and ophthalmic diseases~\cite{chen2015glaucoma}. While CNNs have shown significant promise for analyzing image data, recurrent neural networks (RNN) are better suited for learning long term dependencies in sequential and time-series data~\cite{bengio1994learning}. The state of the art systems in natural language processing (NLP), audio and speech processing use deep RNNs to learn sequential and temporal features ~\cite{salehinejad2017recent}. Finally, hybrid CNN-RNN models have shown significant success in video analytics~\cite{ullah2018action} and speech recognition~\cite{zhao2017recurrent}. These hybrid models show particular promise in cases where both spatial and temporal/sequential features need to be learned from the data. 

Since deep learning came into prominence, it is also being used by researchers for audio based biomedical diagnosis and anomaly detection. Some significant areas of audio based diagnosis using deep learning include sleep apnea detection, cough sound identification, heart sound classification etc. Amoh \emph{et al.}~\cite{amoh2016deep} used a chest mounted sensor to collect audio containing both speech and cough sounds and then used both CNN and RNN to identify cough sounds. Similarly, Nakano \emph{et al.}~\cite{nakano2019tracheal} trained a deep neural networks on tracheal sound spectrograms to detect sleep apnea. In \cite{ryu2016classification}, authors train a CNN architecture to classify heart sounds into normal and abnormal classes. The non-invasive nature of audio based diagnosis make them an attractive choice for biomedical applications.

A major handicap in training a deep network is that a significantly large dataset and considerable time and resources need to be allocated for the training. While the second issue can be solved by using dedicated deep learning accelerators (GPU,TPU etc), the first issue is even more exacerbated for medical research since medical datasets are very sparse and difficult to obtain. One way to circumvent this issue is to use transfer learning. The central idea behind transfer learning is following: a deep network  trained in a domain $D1$ to perform task $T1$ can successfully use the learned data representations to perform task $T2$ in domain $D2$. Most commonly used method for transfer learning is using a large dataset to train a deep network and then re-training a small section of the network on the available data (often significantly small) for the specific task and specific domain. Transfer learning has been used in medical research for cancer diagnosis~\cite{chang2018unsupervised}, prediction of neurological diseases~\cite{payan2015predicting} etc. While traditionally transfer learning refers to transfer of knowledge between two disparate domains, for biomedical research, it is also  used for knowledge transfer in the same domain where a model is trained on a larger population dataset and the knowledge is then transferred for context specific learning on a smaller dataset  \cite{hu2019accurate}\cite{bellot2019boosting}. This strategy is specially useful for biomedical applications due to scarcity of domain specific patient data.

Finally, for employing machine learning methods for medical diagnosis, two primary approaches are used. The first one is generalized models where the models are trained on a database of multiple patient data and it is tested on new patient data. This type of models learns generalized features present across all the patients. While this kind of models are often easier to deploy, they often suffer from inter-patient variability of features and may not produce reliable results for unseen patient data. The second approach is patient specific models, where the models are trained on patient specific data to produce more precise results for the patient specific diagnosis. While these models are harder to train due to difficulty in collecting large amount of patient specific data, they often produce very reliable and consistent results~\cite{kiranyaz2015convolutional}. While a patient specific model requires additional time and effort from healthcare professionals for collecting and labeling the data, specially for chronic diseases where long term monitoring is of the essence, this additional effort is well compensated by reduced hospitalization and reduced loss of time from work for the patient resulting from better continuous monitoring \cite{gibson2000monitoring}. 

Since a large fraction of medical diagnosis algorithms are geared toward wearable devices and mobile platforms, large memory and computational power requirement of deep learning methods present a considerable challenge for commercial deployment. Weight quantization~\cite{zhang2009novel}, low precision computation~\cite{hubara2016binarized} and lightweight networks~\cite{howard2017mobilenets} are some of the approaches used to address this challenge. Quantizing the weights of the trained network is the most straight-forward way to reduce the memory requirement for deployment. DNNs with 8 or 16 bit weights have been shown to achieve comparable accuracy compared to their full precision counterpart~\cite{zhang2009novel}. Though linear quantization is most commonly used, log quantization has been shown to achieve similar accuracy at lower bit precision~\cite{sheng2018quantization}. Finally, lightweight networks like MobileNet~\cite{howard2017mobilenets} reduces computational complexity and memory requirement without significant loss of accuracy  by replacing traditional convolution layers by depthwise separable convolution. 

In this paper we propose a hybrid CNN-RNN model to perform four class classification of breathing sounds on  International  Conference on  Biomedical  and  Health  Informatics  (ICBHI'17)  scientific challenge  respiratory  sound  database~{\cite{rocha2018alpha}} and then devise a screen and model tuning strategy to build patient specific diagnosis models from limited patient data. For comparison of our model with more commonly used CNN architectures, we applied the same methodology on VGGnet~\cite{simonyan2014very} and Mobilenet~\cite{howard2017mobilenets} architecture. Finally, we propose a layerwise logarithmic quantization scheme that can reduce the memory footprint of the networks without significant loss of performance. The sections are organized as follows: section \ref{MandM} describes the dataset, feature extraction method, proposed deep learning model and weight quantization. Section \ref{RandD} tabulates the results for generalized and patient specific model and quantization performance. Finally, section \ref{conclusion} discusses the conclusions and main contributions of the paper.
\section{Materials and Methods}
\label{MandM}
\begin{figure*}[htb]
\centering
\includegraphics[width=0.9\linewidth]{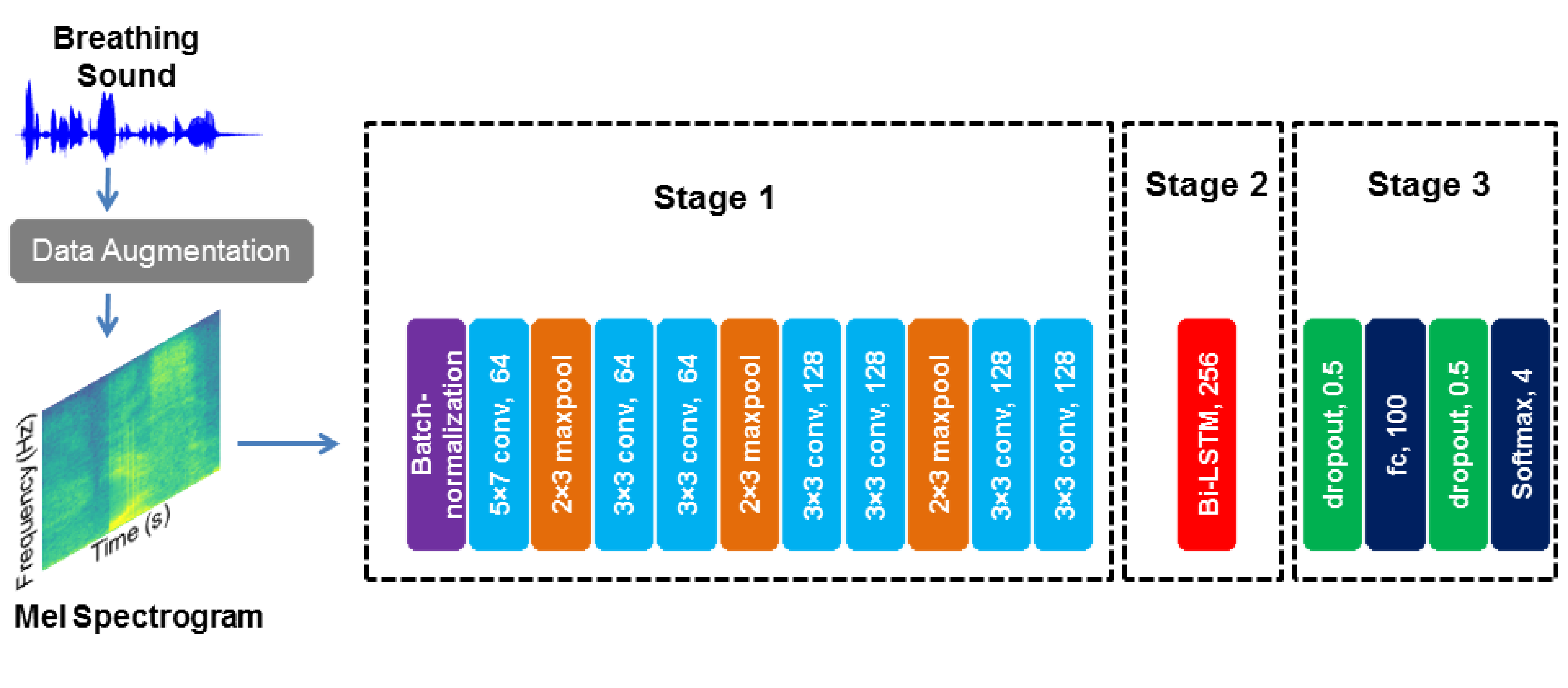}
\caption{Hybrid CNN-RNN: a three stage deep learning model. Stage 1 is a CNN that extracts abstract feature maps from input Mel-spectrograms, stage 2 consists of a Bi-LSTM layer that learns temporal features and stage 3 consists of fully connected (FC) and softmax layers that convert outputs to class predictions.}
\label{fig: fig1}
\end{figure*}
\subsection{Dataset}
For this work we have used the International Conference on Biomedical and Health Informatics (ICBHI'17) scientific challenge respiratory sound database~\cite{rocha2018alpha}. This is the largest publicly available respiratory sound database. The database contains $920$ recordings from $126$ patients. Each breathing cycle in a recording is annotated by respiratory experts as one of the four classes: normal, wheeze, crackle and both (wheeze and crackle). The database contains a total of $6898$ respiratory cycles out of which $1864$ cycles contain crackles, $886$ contain wheeze, $506$ contain both and rest are normal. The dataset contains samples recorded with different equipment (AKG C417L Microphone, 3M Littmann Classic II SE Stethoscope, 3M Litmmann 3200 Electronic Stethoscope and WelchAllyn Meditron Master Elite Electronic Stethoscope) from hospitals in Portugal and Greece. The data is recorded from different locations of chest:      
\begin{inparaenum}[1)]
\item Trachea
\item Anterior left
\item Anterior right
\item Posterior left
\item Posterior right
\item Lateral left and
\item Lateral right.
\end{inparaenum}
 Furthermore, a significant number of samples are noisy. These characteristics make the classification problem more challenging and much closer to real world scenarios compared to manually curated datasets recorded under ideal conditions. Further details about the database and data collection methods can be found in ~\cite{rocha2018alpha}.
\subsection{Evaluation Metrics}
\label{EM}
In the original challenge, out of $920$ recordings, $539$ recordings were marked as training samples and $381$ recordings were marked as testing samples. There are no common patients between training and testing set.  For this work we used the officially described evaluation metrics for the four-class (normal(N), crackle(C), wheeze(W) and both(B)) classification problem defined as follows:
\begin{equation}
    Sensitivity (Se)=\frac{C_{correct}+W_{correct}+B_{correct}}{C_{total}+W_{total}+B_{total}}
\end{equation}
\begin{equation}
    Specificity (Sp)=\frac{N_{correct}}{N_{total}}
\end{equation}

\begin{equation}
    Score (Sc)=\frac{Se+Sp}{2}
\end{equation}
where $i_{correct}$ and $i_{total}$ represent correctly classified and total breathing cycles of $class_i$ respectively. Since deep learning models require a large amount of data for training, we use a 80-20 split of patients for training and testing in all our experiments.The training set contains recordings from 101 patients while testing set contains recordings from 25 patients.

For a more complete evaluation of the proposed model, we also evaluate it using other commonly used metrics such as precision, recall and f1-score. Moreover, the dataset has disproportionate number of normal vs anomalous samples and the official metrics are micro-averaged (calculated over all the classes).Therefore, there is a chance that performance of the models on one class overshadows the other classes in overall results. Therefore, we calculated the precision, recall and f1-score using macro-averaging (metrics are computed for each class individually  and then averaged).
\subsection{Related Work}
A number of papers have been published so far analyzing this dataset. Jakovljevic \emph{et al}~\cite{jakovljevic2018hidden}  used hidden markov model with Gaussian mixture model to classify the breathing cycles. They have used spectral subtraction based noise suppression to pre-process the data and MFCC features are used for classification. Their models obtained a score of $39.56\%$ on the original train-test split and $49.5\%$ on 10-fold cross-validation of the training set. 

Kochetov \emph{et al.}~\cite{kochetov2018noise} proposed a noise marking RNN for the four-class classification. Their proposed model contains two sections: an attention network for binary classification of respiratory cycles into noisy and non-noisy classes and an RNN for four class classification. The attention network learns to identify noisy parts of the audio and suppress those sections and passes the filtered audio to the RNN for classifications. With a 80-20 split, they obtained a score of $65.7\%$. They didn't report the score for the original train-test split. Though this method reports relatively higher scores, one primary issue with this method is that there are no noise labels in the metadata of the ICBHI dataset and the paper doesn't mention any method for obtaining these labels. Since there are no known objective methods to measure the noise labels in these type of audio signals, this kind of manual labeling of the respiratory cycles makes their results unreliable and irreproducible.

Perna \emph{et al.}~\cite{perna2018convolutional} used a deep CNN architecture to classify the breathing cycles into healthy and unhealthy and obtained an accuracy of $83\%$ using a 80-20 train-test split and MFCC features. They also did a ternary classification of the recordings into healthy,chronic and non-chronic diseases and obtained an accuracy of $82\%$.

Chen \emph{et al.}~\cite{chen2019triple} used optimized S-transform based feature maps along with deep residual nets (ResNets) on a smaller subset of the dataset (489 recordings) to classify the samples (not individual breathing cycles) into three classes (N, C and W) and obtained an accuracy of $98.79\%$ on a 70-30 train-test split.

Finally, Chambres \emph{et al.}~\cite{chambres2018automatic} have proposed a patient level model where they classify the individual breathing cycles into one of the four classes using lowlevel features (melbands, mfcc, etc), rythm features (loudness, bpm etc), the SFX features (harmonicity and inharmonicity information) and the tonal features (chords strength, tuning frequency etc). They used boosted tree method for the classification. Next, they classified the patients as healthy or unhealthy based on the percentage of breathing cycles of the patient classified as abnormal. They have obtained an accuracy of $49.63\%$ on the breathing cycle level classification and an accuracy of $85\%$ on patient level classification. The justification for this patient level model is that medical professionals do not take decisions about patients based on individual breathing cycles but rather based on longer breathing sound segments and the trends represented by several breathing cycles of a patient can provide a more consistent diagnosis. A summary of the literature is presented in table \ref{table:1}.

\begin{table*}[!htb]
\caption{Summary of existing literature on ICBHI dataset}
\centering
\begin{tabular}{|>{\centering\arraybackslash}m{2cm}|>{\centering\arraybackslash}m{2cm}|>{\centering\arraybackslash}m{3cm}|m{8cm}|}
\hline
\textbf{Paper} & \textbf{Features} & \textbf{Classification Method} & \multicolumn{1}{|c|}{\textbf{Results}} \\
\hline
Jakovljevic \emph{et al.}~\cite{jakovljevic2018hidden} & MFCC & GMM+ HMM & sc: $39.56\%$ (original train test split), $49.5\%$ (training data, $10$-fold cross-validation)  \\
\hline
Kochetov \emph{et al.}~\cite{kochetov2018noise} & MFCC & Noise masking RNN & sc: $65.7\%$ ($80-20$ split, four class classification)  \\
\hline
Perna \emph{et al.}~\cite{perna2018convolutional} & MFCC & CNN & Acc: $83\%$ ($80-20$ split, healthy-unhealthy classification), Acc: $82\%$ (healthy, chronic and non-chronic classification)  \\
\hline
Chen \emph{et al.}~\cite{chen2019triple} & optimized S-transform &ResNets & sc: $98.79\%$ (smaller subset of original data, $70-30$ split, sample level classification)  \\
\hline
Chambres \emph{et al.}~\cite{chambres2018automatic} & Multiple features & boosted tree & sc: $49.63\%$ (original train test split), Acc: $85\%$ (original train test split, patient level healthy-unhealthy classification)  \\
\hline
\end{tabular}
\vspace{1pt}
\label{table:1}
\end{table*}

\subsection{Proposed Method}
\subsubsection{Feature Extraction and Data augmentation}
Since the audio samples in the dataset had different sampling frequencies, first all of the signals were downsampled to $4kHz$. Since both wheeze and crackle signals are typically present within frequency range $0-2kHz$, downsampling the audio samples to $4kHz$ should not cause any loss of relevant information.

As the dataset is relatively small for training a deep learning model, we used several data augmentation techniques to increase the size of the dataset. We used noise addition, speed variation, random shifting, pitch shift etc to create augmented samples. Aside from increasing the dataset size, these data augmentation methods also help the network learn useful data representations in-spite of different recording conditions, different equipments, patient age and gender, inter-patient variability of breathing rate etc.

For feature extraction we have used Mel-frequency spectrogram with a window size of $60$ ms with $50\%$ overlap. Each breathing cycle is converted to a 2D image where rows correspond to frequencies in Mel scale and columns correspond to time (window) and each value represent log amplitude value of the signal corresponding to that frequency and time window.
\subsubsection{Hybrid CNN-RNN}
\label{HCRNN}
We propose a hybrid CNN-RNN model (figure \ref{fig: fig1}) that consists of three stages: the first stage is a deep CNN model that extracts abstract feature representations from the input data, the second stage consists of a bidirectional long short term memory layer (Bi-LSTM) that learns temporal relations and finally in the third stage we have fully connected and softmax layers that convert the output of previous layers to class prediction. While these type of hybrid CNN-RNN architectures have been more commonly used in sound event detection ({\cite{cakir2017convolutional}}, {\cite{sang2018convolutional}}), due to sporadic nature of wheeze and crackle as well as their temporal and frequency variance, similar hybrid architectures may prove useful for lung sound classification.

The first stage consists of batch-normalization, convolution and max-pool layers. The batch normalization layer scales the input images over each batch to stabilize the training. In the 2D convolution layer the input is convolved with 2D kernels to produce abstract feature maps. Each convolution layer is followed by Rectified Linear activation functions (ReLU). The max-pool layer selects the maximum values from a pixel neighborhood which reduces the overall network parameters and results in shift-invariance ~\cite{litjens2017survey}.

LSTM have been proposed by  Hochreiter and Schmidhuber ~\cite{hochreiter1997long} consisting of gated recurrent cells that block or pass the data in a sequence or time series by learning the perceived importance of data points. Each current output and the hidden state of a cell is a function of current as well as past values of the data. Bidirectional LSTM consists of two interconnected LSTM layers, one of which operates on the same direction as data sequence while the other operates on the reverse direction. So, the current output of the Bi-LSTM layer is function of current, past and future values of the data. We used tanh as non-linear activation function for this layer.

The final fully connected and softmax layers take the output of the Bi-LSTM layer and convert it to class probabilities $p_{class} \in [0,1] $. Finally, the model is trained with categorical crossentropy loss and Adam optimizer for the four class classification problem. We also used dropout regularization in the fully connected layer to reduce overfitting. The parameters for model architecture (such as number of layers, number of filters etc.) were optimized based on minimizing training loss. Finally, to handle the class imbalance pproblem in the dataset, higher costs were assigned to classes with smaller number of samples during model training.

To benchmark the performance of our proposed model, we compare it to two standard CNN models, VGG-16~\cite{simonyan2014very} and Mobilenet~\cite{howard2017mobilenets}. Since our dataset size is limited even after data augmentation, it can cause overfitting if we train these models from scratch on our dataset. Hence, we used Imagenet trained weights instead and replaced the dense layers of these models with an architecture similar to the fully connected and softmax layers of our proposed CNN-RNN architecture. Then the models are trained with a small learning rate.
\subsubsection{Patient Specific Model Tuning}
\label{PSTL}
\begin{figure}[!htb]
\centering
\includegraphics[width=\linewidth]{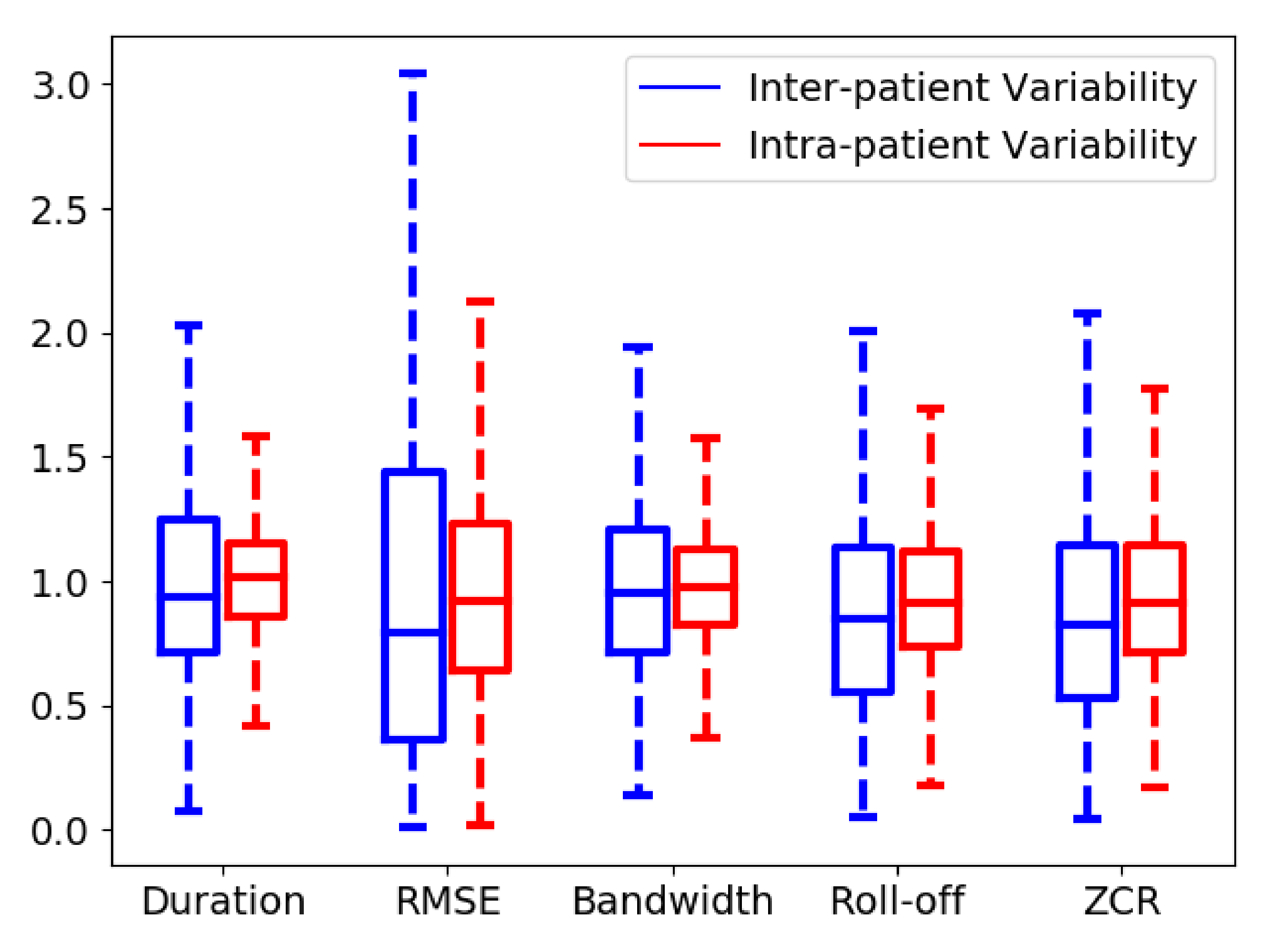}
\caption{Boxplot of intra-patient and inter-patient variability of audio features: Intra-patient variability is computed by normalizing each audio feature by average of that feature for the corresponding patient while for Inter-patient variability, the normalization is done by average of the audio feature over the entire dataset. Diverse set of features are used for comparison including breathig cycle duration, energy related feature (RMS ennergy), noise related feature (ZCR) and spectral features (bandwidth, roll-off). Inter-patient variability is significantly   larger than intra-patient variability for all the cases.}
\label{fig: fig2}
\end{figure}

Though most of the existing research concentrate on developing generalized models for classifying respiratory anomalies, the problem with this kind of models is that their performance can often deteriorate for a completely new patient due to inter-patient variability. This kind of inconsistent performance of classification models make them unreliable and thus hinders their wide scale adoption. To qualitatively    evaluate the inter-patient variability, we show the boxplot of inter-patient and intra-patient variability of a diverse set of audio features  (duration, RMSE, bandwidth, Roll-off, ZCR) in fig. {\ref{fig: fig2}}. For the intra-patient variability, we normalized each audio feature of a sample by average of that feature for the samples from that specific patient while for inter-patient variability, we normalized the audio  features by the average of that feature over the entire dataset. As evident from the figure, inter-patient variability is significantly larger when compared to intra-patient variability.

Also, from a medical professional's perspective, for most of the chronic respiratory patients, some patient data is already available or can be collected and automated long-term monitoring of patient condition after initial treatment is often very important. Though training a model based on existing patient specific data to extract patient specific features result in a more consistent and reliable patient specific model, it is often very difficult to collect enough data from a patient to sufficiently train a machine learning model. Since deep learning models require much larger amount of data for training, the issue is further exacerbated.

To address these shortcomings of existing methods, we propose a patient specific model tuning  strategy that can take advantage of deep learning techniques even with small amount of patient data available. In this proposed model, the deep network is first trained on a large database to learn domain specific feature representations. Then a smaller part of the network is re-trained on the small amount of patient specific data available. This enables us to transfer the learned domain specific knowledge of the deep network to patient specific models and thus produce consistent patient specific class predictions with high accuracy. In our proposed model we train the 3 stage network on the training samples. Then, for a new patient, only the last  stage is re-trained with patient specific breathing cycles while the learned CNN-RNN stage weights are frozen in their pre-trained values. For our proposed strategy only $\sim 1.4\%$ of the network parameters are retrained for patient specific models. For VGG-16 and MobileNet, the same strategy is applied.
\subsubsection{Weight Quantization}
\label{weightquant}
In this proposed weight quantization scheme, the magnitude of weights of each layer are quantized in log domain. The quantized weight ($\Tilde{w}$) can be represented as:
\begin{equation}
    \Tilde{w}=\lfloor 2^N\times \frac{w_{log}-w_{log}^{min}}{w_{log}^{max}-w_{log}^{min}}\rceil
\end{equation}
where $w_{log}$ represents the weights ($w$) mapped to log domain ($\log_{10}(w)$) and N is the bit precision. The total number of bits required to store each weight in this scheme is (N + 1) since one bit is required to store the sign of the bit. Now, the minimum and maximum weights ($w_{log}^{min}$ and $w_{log}^{max}$ ) used for normalization can be calculated globally (over the entire network) or locally (for each layer). Since the architectures used here have different types of layers (convolution, batchnormalization, LSTM etc.) which often show different ranges of weights~\cite{sheng2018quantization}, local weight normaliztion seems to be more logical choice. While local normalization requires minimum and maximum weights of each layer to be saved in memory for retrieving the actual weights, this is very insignificant compared to total memory required to save the quantized weights. Finally, we rounded very small weights to zero before applying log quantization to limit the quantization range in log domain.

\section{Results and Discussions}
\label{RandD}
\subsection{Generalized Model}
\label{GM}
\begin{table*}[htb]
\centering
\caption{Comparison of results}
\begin{tabular}{|>{\centering\arraybackslash}m{2cm}|>{\centering\arraybackslash}m{1.5cm}|>{\centering\arraybackslash}m{1.5cm}|>{\centering\arraybackslash}m{1.5cm}|>{\centering\arraybackslash}m{1.5cm}|>{\centering\arraybackslash}m{1.5cm}|>{\centering\arraybackslash}m{1.5cm}|}
\hline
\multirow{2}*{\textbf{Model}} & \multicolumn{3}{|c|}{\textbf{Micro Metrics}} & \multicolumn{3}{|c|}{\textbf{Macro Metrics}} \\
\cline{2-7}
& \textbf{Sensitivity} & \textbf{Specificity} & \textbf{Score} &\textbf{Precision} & \textbf{Recall} & \textbf{f1-score} \\
\hline
VGG-16 & $40.63\%$ & $86.03\%$ & $63.33\%$ & $49.09\%$ & $49.24\%$ & $48.99\%$ \\
\hline

Mobilenet & $46.31\%$ & $78.20\%$ & $62.26\%$ & $52.02\%$ & $55.21\%$ & $53.13\%$ \\
\hline
 Our work & $48.63\%$  & $84.14\%$ & $\textcolor{red}{66.38\%}$ & $58.47\%$  & $58.01\%$ & \textcolor{red}{$57.91\%$}\\
 \hline
\end{tabular}
\label{table:2}
\end{table*}

Firstly, we evaluated our model on four class breathing cycle  using micro and macro metrics described in section \ref{EM}. To compare our results with traditionally used CNN architectures, we used VGGnet and Mobilenet. All models were trained and tested on a workstation with Intel Xeon E5-2630 CPU with 128GB RAM and NVIDIA TITAN Xp GPU. The results are tabulated in table \ref{table:2}. The results are averaged over five randomized train-test sets. As it can be seen from the table \ref{table:2}, the proposed hybrid CNN-RNN model trained with data augmentation produces state of the art results. Both VGG-16 and Mobilenet produces slightly lower scores both in terms of macro and micro metrics. The score obtained by the proposed model also outperform results reported by Kochetov \emph{et al.} using noise labels on a similar 80-20 split (Table \ref{table:1}). 
We have also performed a 10-fold cross-validation on the dataset for our proposed model and the average score obtained is $66.43\%$.
Due to unavailability of similar audio datasets in biomedical field, we have also tested the proposed hybrid model on Tensorflow speech recognition challenge {\cite{warden2017speech}}  to benchmark its performance. For an eleven-class classification with $90\%-10\%$ train-test split, it produced a respectable accuracy of $~96\%$.
For the sake of completeness, we also tested the dataset using same train-test split strategy with a variety of commonly used temporal and spectral features (RMSE, ZCR, spectral centroid, roll-off frequency, entropy, spectral contrast etc. {\cite{pramono2019evaluation}}) with non-DL methods such as SVM, shallow neural network, random forest and gradient boosting. The resulting scores were significantly lower ($44.5-51.2\%$).
\subsection{Patient Specific Model Tuning Strategy}
\label{PSTLM}
\begin{figure*}[!htb]
\centering
\includegraphics[width=\linewidth]{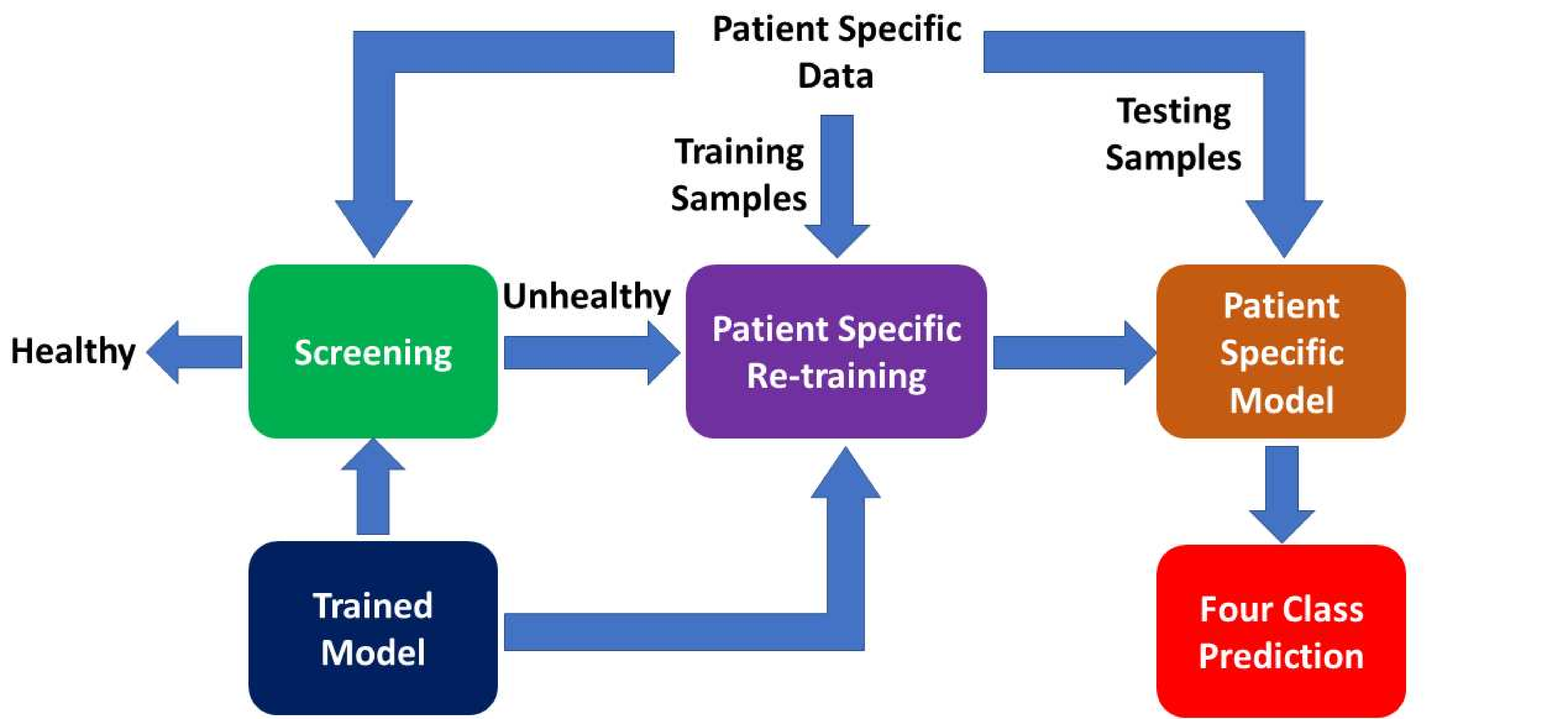}
\caption{Screen and model tuning strategy: First the patients are screened into healthy and unhealthy based on $\%$ of breathing cycles predicted as unhealthy. For patients predicted to be unhealthy, trained model is re-trained on patient specific data to produce patient specific model which then performs the four class prediction on breathing cycles.}
\label{fig: fig3}
\end{figure*}

It has been shown by Chambres \emph{et al.}~\cite{chambres2018automatic} that though it is difficult to achieve high scores for breathing cycle level classification, it is much easier to achieve high accuracy in patient level binary classification (healthy/sick). Hence, we propose a screen and model tuning strategy. First, the patients are screened using the pre-trained model and if a patient is found to be unhealthy, the pre-trained model is retrained on the patient data to build the patient specific model that can monitor the patient condition in future with higher reliability. The proposed model is shown in Fig. \ref{fig: fig3}. To evaluate the performance of the proposed methodology, we used leave one out validation. Since there are variable number of recordings from each patient in the dataset, of the $n$ samples from a patient, $n-1$ samples are used to retrain the model and it is tested on the other sample. This method is repeated so that all the samples are in the test set once. We trained the proposed model on the patients in the train set and evaluated it on the patients in test set. Since leave one out validation is not possible for patients with only one sample, we only considered patients with more than one sample.
The dataset contains different number of recordings from each patient and the length of the recordings and number of breathing cycles in each recording vary widely. But on average, $\approx 47$ patient breathing cycles are used for the fine-tuning of patient specific models.

\begin{table*}[htb]
\centering
\caption{Comparison of patient specific Models}
\begin{tabular}{|>{\centering\arraybackslash}m{6cm}|>{\centering\arraybackslash}m{3cm}|>{\centering\arraybackslash}m{1.5cm}|>{\centering\arraybackslash}m{1.5cm}|>{\centering\arraybackslash}m{1.5cm}|}
\hline
\textbf{Pre-trained Model} & \multicolumn{1}{|c|}{\textbf{Patient Specific Training}} & \multicolumn{1}{|c|}{\textbf{Se}} & \multicolumn{1}{|c|}{\textbf{Sp}}& \multicolumn{1}{|c|}{\textbf{Sc}} \\
\hline
NA. & Majority Class & $46.25\%$ & $80.45\%$ & $63.35\%$\\
\hline
VGG-16 (ImageNet) & SVM & $50.35\%$ & $82.38\%$ & $66.36\%$\\
\hline
VGG-16 (ICBHI) & Dense + Softmax & $54.41\%$ & $82.66\%$ & $68.54\%$\\
\hline
MobileNet (ICBHI) & Dense + Softmax & $51.28\%$ & $83.92\%$ & $67.60\%$\\
\hline
Hybrid CNN-RNN (TF speech dataset) & Dense + Softmax & $43.01\%$ & $74.28\%$ & $58.4\%$\\
\hline
Hybrid CNN-RNN (UrbanSound8K dataset) & Dense + Softmax & $34.73\%$ & $88.45\%$ & $61.59\%$\\
\hline
Hybrid CNN-RNN (ICBHI dataset) & Dense + Softmax & $56.91\%$ & $86.70\%$ & $\textcolor{red}{71.81\%}$\\ 
\hline
\end{tabular}
\label{table:3}
\end{table*}

Now, for an objective evaluation of the proposed method, we need to consider a few things. Firstly, if for a patient, most of the breathing cycles belong to one of the classes, the simplest strategy is to assign that class to any of the breathing cycles in the test set. We define this strategy as majority class and use it as a baseline to evaluate the performance of our model. The predicted class probability of $k^{th}$ class is defined as:
\begin{equation}
\begin{split}
    p_{class^k} & = 1 \quad if \quad N_B^k > N_B^i \quad \forall \quad i \neq k \\
                & = 0 \quad otherwise
\end{split}
\end{equation}
where $N_B^i$ is the number of breathing cycles belonging to class $i$ to the specific patient.

Secondly, since we are using patient specific data to train the models, we have to verify if our proposed model tuning strategy provides any advantage over a simple classifier trained on only patient specific data. To verify this, we used an ImageNet~\cite{deng2009imagenet} trained  VGG-16~\cite{simonyan2014very} as a feature extractor along with an SVM classifier to build patient specific models. Variants of VGG trained on ImageNet dataset have been shown to be very efficient feature extractors not only for image classification, but also for audio classification~\cite{hershey2017cnn}. Here we use the pre-trained CNN to extract features from patient recordings and train an SVM based on those features only on the patient specific data.

Thirdly, we are proposing that by pre-training the hybrid CNN-RNN model on the respiratory data, the model learns domain specific feature representations that are transferred to the patient specific model. To justify this claim, we trained the same model on tensorflow speech recognition challenge dataset  \cite{warden2017speech} as well as urban sounds 8K dataset \cite{urbansound}. Then we used the same model tuning strategy to re-train the model on patient specific data. If the proposed model learns only the audio feature specific abstract representations from the data, then a model trained on any sufficiently large audio database should perform well. But, if the model learns respiratory sound domain specific features from the data, the model pre-trained on respiratory sounds should outperform the model pre-trained on any other type of audio database. Finally, we comapare the results of our model with pure CNN models VGG-16 and MobileNet using the same experimental methodology. 

The results are tabulated in table~\ref{table:3}. Firstly, Our proposed strategy outperforms all other models and strategies and obtains a score of $71.81\%$. Secorndly, VGG-16  and MobileNet achieves scores $68.54\%$ and $67.60\%$ which signifies pure  CNNs can be employed for respiratory audio classification, albeit not as effective as a CNN-RNN hybrid model. Thirdly, results corresponding to both audio trained networks shows that audio domain pre-training is not very effective for respiratory domain feature extraction. We explain this observation in further details in section \ref{conclusion}. Finally, Imagenet trained VGG-16 shows promise as a feature extractor for respiratory data, although it does not reach the same level of performance as ICBHI trained models.
\subsection{Memory and Computational Complexity}
\label{MCC}
\begin{figure}[!htb]
\centering
\includegraphics[width=\linewidth]{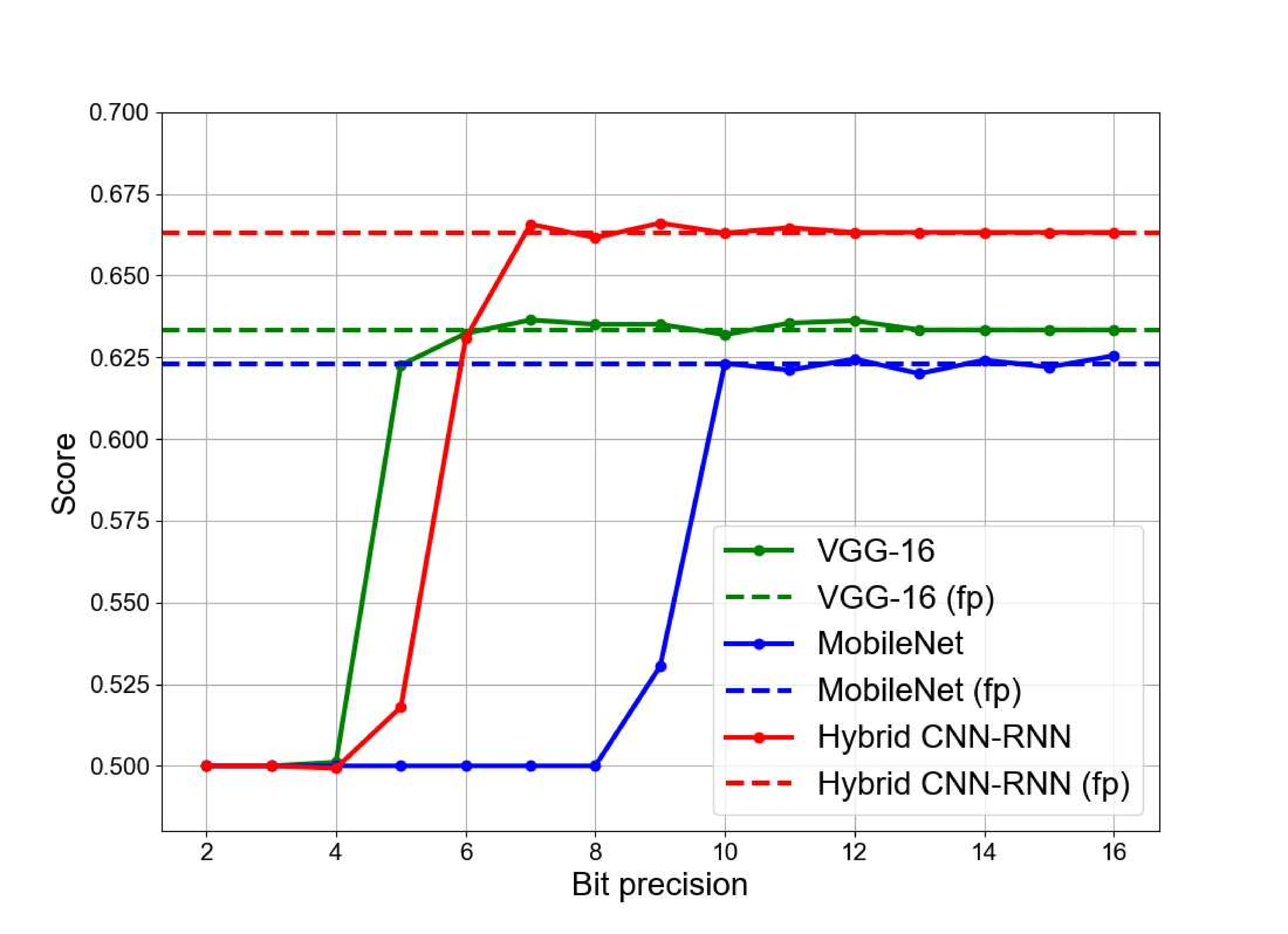}
\caption{Local log quantization: Score achieved by VGG-16, MobileNet and hybrid CNN-RNN with varying bit precision under local log quantization. VGG-16 requires minimum bit precision to achieve full precision (fp) accuracy while MobileNet requires maximum bit precision.}
\label{fig: fig4}
\end{figure}
Even though the proposed models show excellent performance in the classification task, the memory requirement for storing huge number of weights for these models make them unsustainable for application in mobile and wearable platforms. Hence, we apply the local log quantization scheme proposed in section \ref{weightquant}. Figure \ref{fig: fig4} shows the score achieved by the models as a function of bit precision of weights. As expected, VGG-16 outperforms the other to models due to its over-parameterized design~\cite{sheng2018quantization}. MobileNet shows particularly poor performance in weight quantization and is only able to achieve optimum accuracy at $10$ bit precision. This poor quantization performance can be attributed to large number of batch-normalization layers and RELU6 activation of MobileNet architecture~\cite{sheng2018quantization}. While several approaches have been proposed to circumvent these issues~\cite{alyamkin2019low}, these methods are not compatible with Imagenet pre-trained  MobileNet model since they focus on modifications in the architecture rather than quantization of pre-trained weights. The hybrid CNN-RNN model performs slightly worse than VGG-16 since it has LSTM layer which requires higher bit precision compared to the CNN counterpart~\cite{gokmen2018training}.

\begin{figure}[!htb]
\centering
\includegraphics[width=\linewidth]{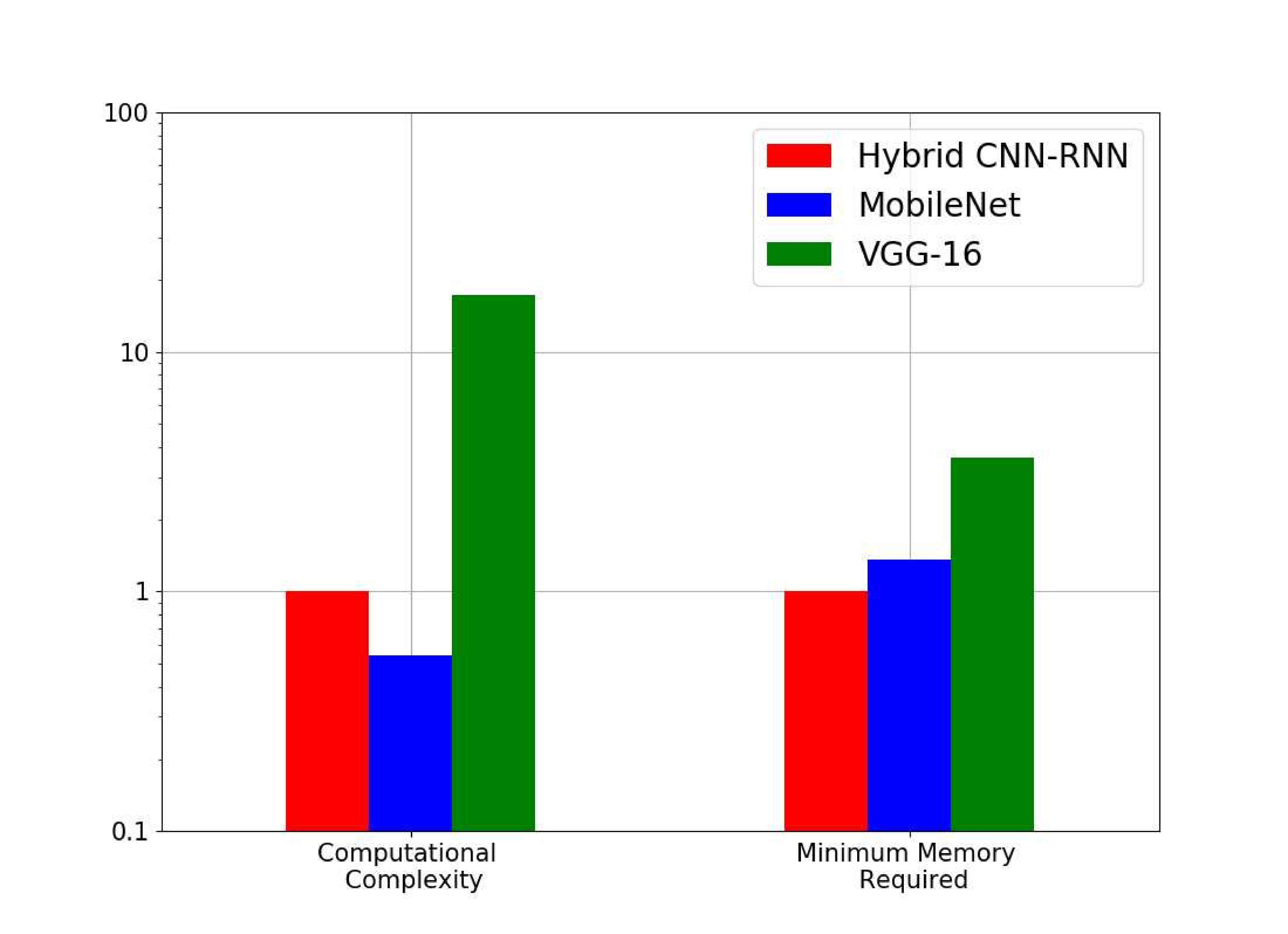}
\caption{Resource comparison: Comparison of normalized computational complexity (GFLOPS/sample) and minimum memory required (Mbits) by VGG-16, MobileNet and hybrid CNN-RNN. MobileNet and hybrid CNN-RNN present a trade-off between computational complexity and memory required for optimum performance.}
\label{fig: fig5}
\end{figure}

Finally, we compare the computational complexity and minimum memory requirement of the models. The computational complexity is calculated in GFLOPS per sample during inference. The minimum memory required for a model is calculated as (total number of parameters $\times N^{opt}$) where $N^{opt}$ is the minimum bit precision required for the model to achieve optimum accuracy. The only other major source of computational complexity for end-to-end classification of respiratory audio is computations required for converting audio samples to their respective mel-spectrograms. Since this compuatational overhead for feature extraction is significantly less ($<1$ MFLOPS)~{\cite{acharya2017feature}} compared to the DL architecture, we can ignore it in computational complexity calculations. Both computational complexity and memory are normalized with respect to hybrid CNN-RNN model and shown in figure \ref{fig: fig5}. Due to large number of parameters, even after significant memory compression through weight quantization, VGG-16 requires significantly higher memory. Hybrid CNN-RNN model requires least amount of memory followed by MobileNet. In terms of computational complexity, while VGG-16 requires very high computational overhead, MobileNet is computationally most efficient followed by hybrid CNN-RNN. Hence, MobileNet is more suitable for a hardware system where power is the primary constraint while hybrid CNN-RNN, while less power efficient than MobileNet, achieves better performance at minimal memory footprint.

Finally, Our proposed system requires data pre-processing, feature extraction and classification only once in each breathing cycle. Therefore, if we consider a ping-pong buffer architecture {\cite{katz_2007}} for audio acquisition and processing, our system needs to perform end to end classification of breathing cycles at a latency smaller than minimum breathing cycle duration for real time operation. The primary computational bottleneck of the proposed system is the DL architecture as mentioned earlier. The number of computations of the proposed architecture is of the same order as Mobilenet as shown in fig. {\ref{fig: fig5}}. Since the minimum breathing cycle duration is $>1$ second {\cite{lindh2013delmar}} and 
the per sample latency of Mobilenet on modern mobile SoCs is only $\sim100$ ms {\cite{ignatov2018ai}}, the proposed system should easily be able to perform real time classification of respiratory anomalies.

\section{Conclusion}
\label{conclusion}
In this paper, we have developed a hybrid CNN-RNN model that produces state of the art results for ICBHI'17 respiratory audio dataset. It produces a score of $66.31\%$ score on 80-20 split for four-class respiratory cycle classification. We also propose a patient screening and model tuning strategy to identify unhealthy patients and then build patient specific models through patient speecific re-training. This proposed model provides significantly more reliable results for the original train-test split achieving a score of $71.81\%$ for leave-one-out cross-validation. It is observed that trained models from image recognition field, surprisingly, perform better in transferring knowledge than those pre-trained on speech. A possible explanation for this could be that while image-trained models are trained on a much larger Imagenet dataset and therefore, has better generalization performance compared to models trained on relatively smaller audio datasets. While lack of availability of pre-trained models in audio domain and prohibitively long training time required for training a model with audio datasets of sizes comparable to Imagenet prevent us from verifying this hypothesis in this work, in future we plan to explore transfer learning performance of audio  and image datasets in further detail. We also develop a local log quantization strategy for reducing the memory cost of the models that achieves $\approx 4\times$ reduction in minimum memory required without loss of performance. The primary significance of this result is that this weight quantization strategy is able to achieve considerable weight compression without any architectural modification to the model or quantization aware training. Finally, while the proposed model has higher computational complexity than MobileNet, it has minimal memory footprint among the models under consideration. Since the amount of data from a single patient is still very small for this dataset, in future, we plan to employ this strategy with larger amount of patient specific data. We also plan to create an embedded implementation of this algorithm for a wearable device to be used in patient monitoring at home. Further reductions in computational complexity will be explored using a neuromorphic spike based approach~\cite{jetcas_review,Jyoti-Shih-Frontiers}. 
\FloatBarrier

\ifCLASSOPTIONcaptionsoff
  \newpage
\fi

\bibliographystyle{IEEEtran}
\bibliography{main.bib}

\end{document}